# *In-situ* fabrication of cobalt-doped SrFe$_2$As$_2$ thin films by using pulsed laser deposition with excimer laser

Eun-Mi Choi,[a] Soon-Gil Jung,[a] Nam Hoon Lee,[a] Young-Seung Kwon,[a] Won Nam Kang,[a,*] Dong Ho Kim,[b] Myung-Hwa Jung,[c] Sung-Ik Lee,[c] and Liling Sun[d]

[a] BK21 Physics Division and Department of Physics, Sungkyunkwan University, Suwon 440-746, Republic of Korea

[b] Department of Physics, Yeungnam University, Gyeongsan 712-749, Republic of Korea

[c] Department of Physics, Sogang University, Seoul 121-742, Republic of Korea

[d] National Laboratory for Superconductivity, Institute of Physics, Chinese Academy of Sciences, Beijing 100190, P. R. China

Abstract

The remarkably high superconducting transition temperature and upper critical field of iron(Fe)-based layered superconductors, despite ferromagnetic material base, open the prospect for superconducting electronics. However, success in superconducting electronics has been limited because of difficulties in fabricating high-quality thin films. We report the growth of high-quality *c*-axis-oriented cobalt(Co)-doped SrFe$_2$As$_2$ thin films with bulk superconductivity by using an *in-situ* pulsed laser deposition technique with a 248-nm-wavelength KrF excimer laser and an arsenic(As)-rich phase target. The temperature and field dependences of the magnetization showing strong diamagnetism and transport critical current density with superior $J_c - H$ performance are reported. These results provide necessary information for practical applications of Fe-based superconductors.

\pacs{}

The classes of Iron (Fe)-based superconductors [1-8] have remarkably high superconducting transition temperatures ($T_c$) in spite of the ferromagnetic material base; the highest $T_c$ is 55 K in LnFeAsO (FeAs-1111, Ln = lanthanide) [2] and 37.5 K in $AE$Fe$_2$As$_2$ (FeAs-122, $AE$ = alkaline-earth element) [7]. The zero-temperature upper critical field ($H_{c2}(0)$) was found to be up to 65 T in the FeAs-1111 [9]. The discovery of these new classes of superconductors has regenerated interest in superconductivity because of an opportunity to tune these materials in many ways [10, 11]. This potentially allows one to reveal the mechanism of high-temperature superconductors. Ever since discovering these compounds, much progress has been made in measuring the fundamental physical properties in order to understand the superconducting mechanism. However, controversy still exists because the measurements were carried out on bulk polycrystals, except for some works on single crystals [8, 12-16]. High-quality thin films are needed to investigate the physical properties and to develop superconducting electronic devices, such as Josephson junctions. However, controlling the stoichiometry of the FeAs-1111 phase is difficult because the crystal structure contains two different anions [17]. Also, in the FeAs-1111 phase, electrons are doped by partially replacing oxygen ions with fluorine (F), which is easily evaporated in a vacuum chamber at a high temperature because of its very high vapor pressure. These reasons make it difficult to fabricate high-quality thin film [17-19].

Very recently, Hosono *et al.* reported success in growing cobalt (Co)-doped SrFe$_2$As$_2$ thin film [18, 19]. Even though this compound has a relatively lower $T_c$, Co-doping is more suitable for thin film growth than other types of doping (F or potassium (K)) because of the low vapor pressure of Co. Also, SrFe$_2$As$_2$ contains only one anion species. Hosono *et al.* fabricated the thin films by using pulsed laser deposition (PLD) with a second-harmonic 532-nm-wavelength Nd:YAG laser and a stoichiometric target disk.

In general, an ultraviolet (UV) wavelength is known to be more effective than a visible or an infrared (IR) wavelength in PLD. The wavelength of the laser affects the topology of the target and the size of the particulates [20]. One also has to consider an arsenic (As) deficiency during deposition. The concentration of As plays a very important role in determining the sample's quality. When we used a stoichiometric target in growing thin films under high vacuum and high substrate temperature ($T_s$), we were not able to obtain a high-quality superconducting thin film because of the As deficiency.

To solve this problem, we investigated a suitable As-rich phase target by increasing the amount of As up to 40% and found that a 30% As-rich target was suitable to obtain high-quality films by using PLD with an excimer laser. As grown thin films on LaAlO$_3$ (100) and Al$_2$O$_3$ (0001) substrates by using the PLD with an excimer laser with a 248-nm wavelength and a 30% As-rich target show higher $T_{c,0}$ with a low resistivity and a smooth surface. Also, we report the temperature and field dependences of the magnetization in films showing strong diamagnetism and transport critical current density ($J_c$) with superior $J_c - H$ performance.

To synthesize the target for PLD, elements Sr, Co, As, and FeAs in a stoichiometric molar ratio were put into a quartz tube and then sealed under vacuum. In order to compensate for the As deficiency in the resulting thin film, we added extra As (30%). The more excess amount of As than 30% leads to segregate As from target. The quartz ampoule was heated at 900℃ for 12 h. The resulting material was well ground under a high-purity (99.999%) Ar atmosphere and sintered again at 900℃ for 12 h. The final material was well ground and molded into a disk shape of 3 mm in thickness and then heat treated at 800℃ for 8 h after having been encapsulated in a quartz ampoule. The quality of target is very sensitive to amount of extra As and to the details of heat treatment. The resistance of well-made target is several Ω.

Thin films were fabricated over a wide temperature range from 770℃ to 820℃ at a high vacuum of $10^{-6}$ Torr. In order to remove the residual oxygen in the vacuum chamber, we baked the growth chamber before film depositions by heating the susceptor to a temperature of 600℃ for 20 min under diluted H$_2$ atmosphere. The laser beam was generated by using a Lambda Physik KrF excimer laser with a 248-nm wavelength. The laser's energy density was 1.15 J/cm$^2$, and its frequency was 48 Hz. The growth rate was 3 – 4 nm/s under these deposition conditions.

The phase and the crystalline quality of the thin film were investigated by using X-ray diffractometry (XRD). The thickness and the surface morphology of the thin film were confirmed by using scanning electron microscopy (SEM). The superconducting properties were measured by using a physical property measurement system (PPMS, Quantum Design) and a vibrating sample measurement system (SQUID-VSM, Quantum Design). The current-voltage characteristics were investigated by using an AC four-probe method in various magnetic fields applied parallel and perpendicular to the *c*-axis of the films by using PPMS.

Figure 1a shows X-ray diffraction spectra of thin films deposited on (0001) Al$_2$O$_3$ and (100) LaAlO$_3$. In the case of films grown on LaAlO$_3$, the (00l) peaks of the Co-doped SrFe$_2$As$_2$ are very sharp and strong, and minor impurity phases of FeAs and Fe$_2$As also appear. These impurity phase probably originate from the higher growth temperature [17-19]. The Inset shows a rocking curve of the (008) diffraction peak, and the full width at half maximum (FWHM) of the (008) peak is 0.8°, indicating that thin films grown on Al$_2$O$_3$ and LaAlO$_3$ have highly *c*-axis-oriented crystal structures, despite of the lattice mismatch.

Figure 1b and c show SEM images of interface between the film and the substrate (Fig. 1b), and the surface morphology (Fig. 1c) of the film grown on LaAlO$_3$ at 800℃. Our thin film showed a very clean interface, reflecting no severe chemical reaction between the film and the substrate, and a smooth surface. The thickness of film was typically 700 – 800 nm. The surface morphology showed a well-linked granular structure.

Figure 2 shows the temperature dependences of the resistivity ($\rho$-$T$, Fig. 2a) and of the magnetization ($M$ – $T$, Fig. 2b) for a 700-nm-thick thin film on a LaAlO$_3$ substrate. The onset of $T_c$ is about 20 K, which is the same as that of the polycrystalline bulk samples [19]. The zero-resistance transition temperature, $T_{c,0}$, is observed to be 16.4 K, which is slightly higher than the $T_{c,0}$ recorded in previous reports [18, 19, 21]. The $T_{c,50\%}$, 50% transition of the normal state, is 18.1 K, which is higher by ~ 1 K, and the transition width, $\Delta T_c$ ($T_{c,90\%}$ - $T_{c,10\%}$), is 2.2 K, which is narrower than previous results [18, 21]. The residual resistivity at 30 K is 3.2 × 10$^{-4}$ Ωcm, with a residual resistivity ratio, $RRR = \rho_{300K}/\rho_{30K}$, of 1.7.

The zero-field-cooled magnetizations show a strong diamagnetic signal. These results indicate that our thin film had good bulk superconducting properties, which could be attributed to the high purity of the phase due to the supply of a sufficient excess amount of As. The magnetization hysteresis ($M$ – $H$) curves (inset of Fig. 2b) of the superconducting state at 2 K and of the normal state at 30 K of the thin film were obtained in the present work. The $M$ – $H$ curve at 2 K shows a clear superconducting hysteresis even though a ferromagnetic signal is included while the $M$ – $H$ curve at 30 K shows a formal ferromagnetic hysteresis. These results indicate that superconductivity becomes strong well below $T_c$ even though ferromagnetic impurities exist in the thin film. We consider Fe$_2$As, FeAs, Fe, and Co as possible magnetic impurities. We can infer that these impurities account for the broad transition in the $M$ – $T$ data (Fig. 2b) [22-24]. It should be noticed that Bean's critical state model is not applicable in calculating $J_c$ because of the ferromagnetic signal.

Another main result in the present paper is the transport critical current density ($J_c$), which was measured at 5 (squares) and 10 (triangles) K as functions of the magnetic field parallel ($H$//$ab$, open symbols) or perpendicular ($H$//$c$, solid symbols) to the film's surface, as shown in Fig. 3. These data were measured for a film with a width of 1.6 mm and a thickness of 700 nm and $J_c$ was determined by using an electric-field criterion of 1 $\mu$V/cm. The $J_c$ was observed to be 1.6 × 10$^4$ A/cm$^2$ at 5 K and 1000 Oe, which is relatively very low compared to other superconductors. However, the $J_c$ drops very slowly, just by only 1 order of magnitude, for fields up to 5 T. We could infer that this result is due either to a high density of strong pinning centers caused by a magnetic impurity and a grain boundary or to a high $H_{c2}$. Inset shows the temperature dependence of the upper critical field ($H_{c2}$) for the Co-doped SrFe$_2$As$_2$ thin film. The $H_{c2}$ values are determined from the 90% dropoff of the normal-state resistance at 20 K. The d$H_{c2}^{ab}$/d$T$ and the d$H_{c2}^{c}$/d$T$ for the $H_{c2}$ – $T$ curves are -2.78 and -2.28 *T/K*, respectively. We obtain $H_{c2}^{ab}(0)$ ~ 46 T and $H_{c2}^{c}(0)$ ~ 39 T by using a simple linear extrapolation [21]. The estimated anisotropic ratio at zero temperature, $\gamma = H_{c2}^{ab}(0)/H_{c2}^{c}(0)$, is about 1.2, which is similar to previously reported results [18, 21]. With a high $H_{c2}$, such a superior $J_c$ – $H$ performance would be a very strong advantage for practical applications under higher fields.

In summary, we report the growth of high-quality *c*-axis-oriented Co-doped SrFe$_2$As$_2$ thin films with bulk superconductivity by using an *in-situ* pulsed laser deposition technique with an UV 248-nm-wavelength KrF excimer laser and an 30% As-rich phase target to compensate for the arsenic deficiency in the films. As-grown thin films on LaAlO$_3$ substrates show higher $T_{c,0}$, $T_{c,50\%}$ and sharp $\Delta T_c$ with a low resistivity and a smooth surface. We report the temperature and field dependences of the magnetization in films showing strong diamagnetism and transport critical current density with superior $J_c$ – $H$ performance. Although further refinement is needed, the results presented in this report supply a method to fabricate high-quality *in-situ* thin films of Co-doped SrFe$_2$As$_2$. We believe that the present result will provide a good prospect for fabrication of other Fe-based superconducting thin films and necessary information for their practical applications.

## ACKNOWLEDGMENTS

This work was supported by the Korea Science and Engineering Foundation (KOSEF) grant funded by the Korea government (MEST, No. R01-2008-000-20586-0) and by a grant (R-2006-1-248) from Electric Power Industry Technology Evaluation & Planning (ETEP), Republic of Korea.

**FIGURES**

FIG. 1. (a) X-ray diffraction spectra ($\theta$ - $2\theta$ scan). (b) Scanning electron microscopy images of the interface and the surface morphology of a highly *c*-axis-oriented Co-doped $SrFe_2As_2$ thin film.

FIG. 2. Superconducting properties of a Co-doped $SrFe_2As_2$ thin film grown on a $LaAlO_3$ substrate. (a) $\rho$(T). (b) *M*(T) at *H* = 5 Oe. The inset shows magnetization hysteresis (*M* – *H*) curves in the range of -2000 ≤ *H* ≤ 2000 Oe at 2 K and 30 K.

FIG. 3. The field dependence of $J_c$ when the magnetic field is applied perpendicular (solid symbols) and parallel (open symbols) to the film at *T* = 5 K (squares) and 10 K (triangles) from the *I-V* curves. The inset shows $H_{c2}$(T).

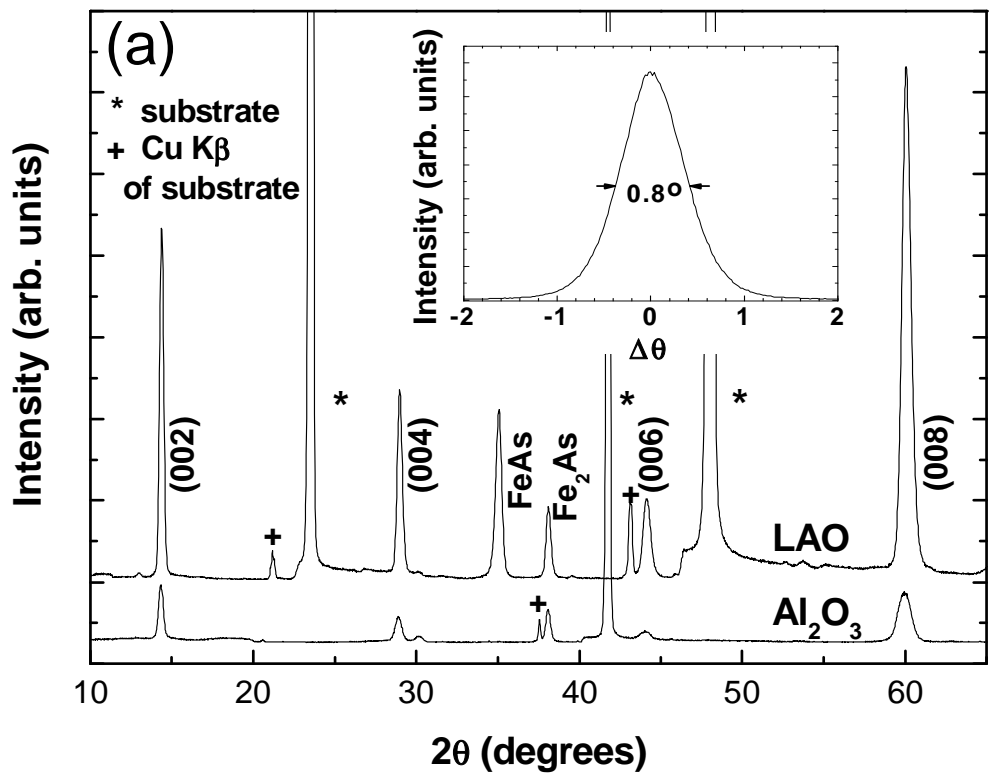
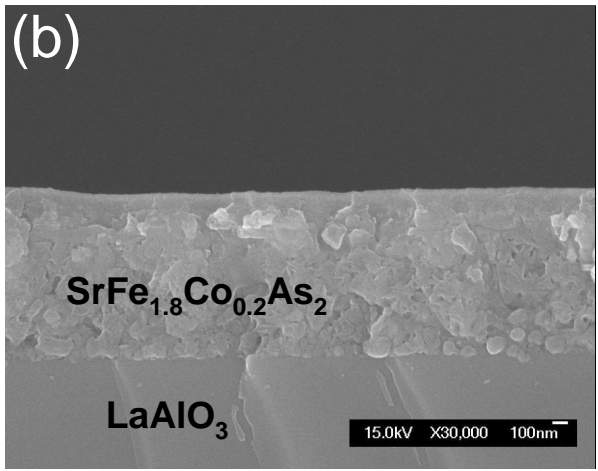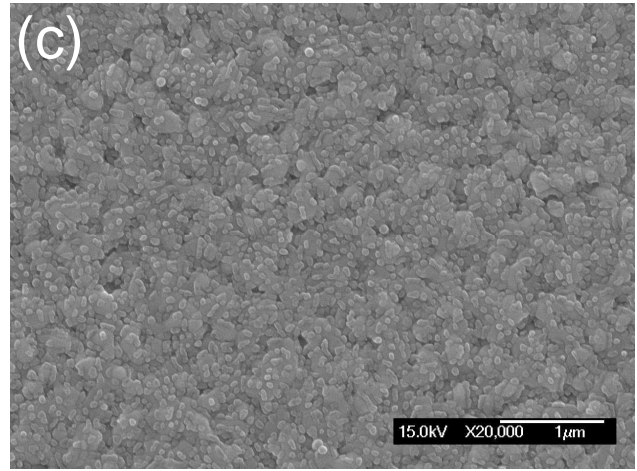

Figure 1 Eun-Mi Choi *et al., In-situ* fabricaion……

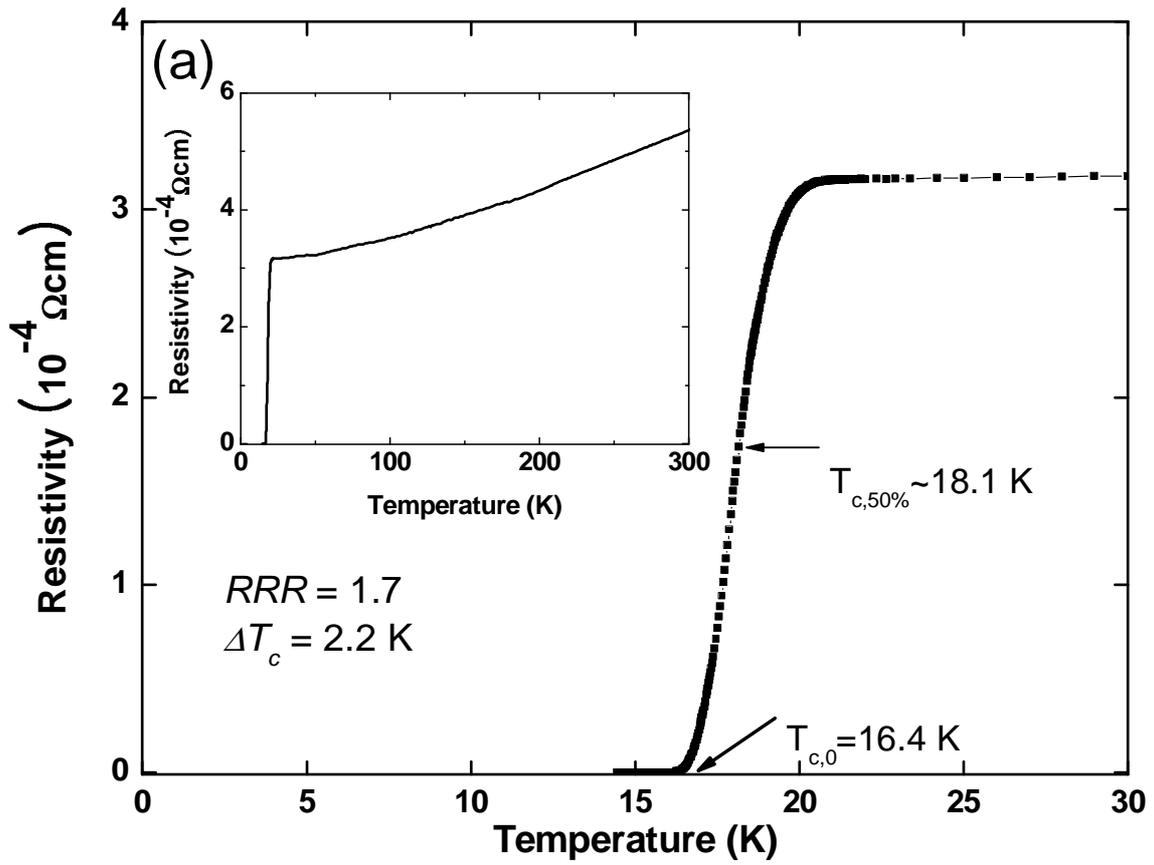

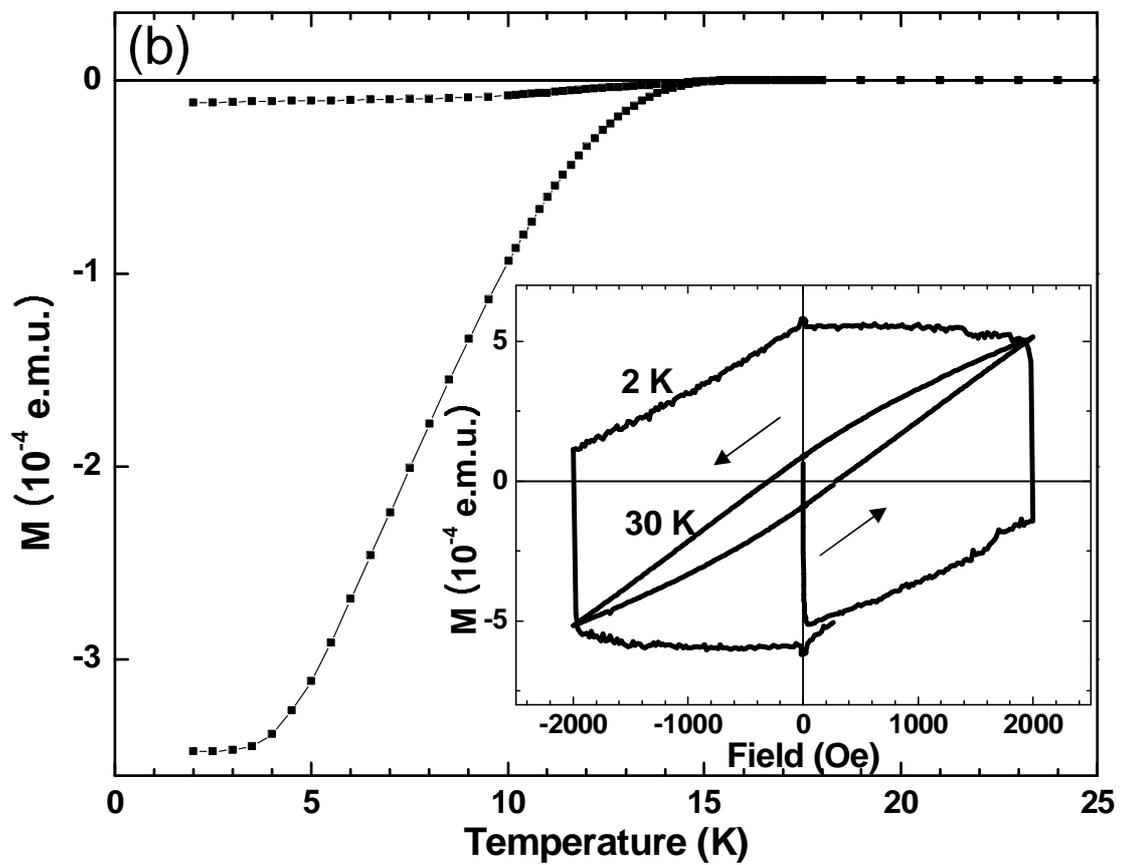

Figure 2 Eun-Mi Choi *et al., In-situ* fabricaion……

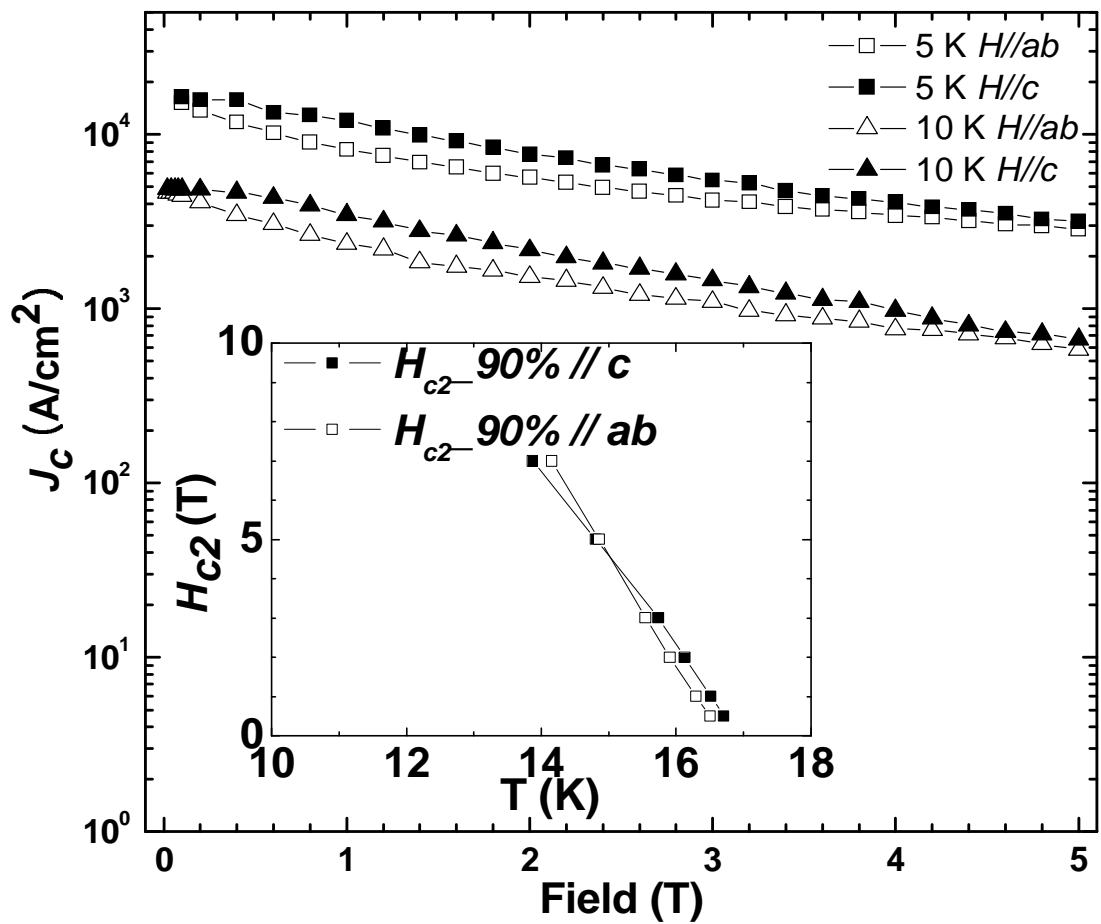

Figure 3 Eun-Mi Choi *et al., In-situ* fabricaion……